\documentclass[preprint]{aastex}

\slugcomment{}

\shorttitle{Infant Super Star Clusters}
\shortauthors{}

\begin{document}

\title{The Spectral Energy Distributions of Infant Super Star Clusters 
in Henize 2-10 from 7~mm to 6~cm}


\author{Kelsey E. Johnson \altaffilmark{1}}
\affil{Department of Astronomy, University of Wisconsin,
    Madison, WI 53706 \\ and \\ 
National Radio Astronomy Observatory, Socorro, NM 87801} 
\and

\author{Henry A. Kobulnicky}
\affil{Department of Physics and Astronomy, University of Wyoming, 
Laramie, WY 82071}


\altaffiltext{1}{NSF Astronomy \& Astrophysics Postdoctoral Fellow}

\begin{abstract}
We present observations from our continuing studies of the earliest
stages of massive star cluster evolution.  In this paper, radio
observations from the Very Large Array at 0.7~cm, 1.3~cm, 2~cm,
3.6~cm, and 6~cm are used to map the radio spectral energy
distributions and model the physical properties of the ultra-young
embedded super star clusters in Henize~2-10.  The 0.7~cm flux
densities indicate that the young embedded star clusters that are
powering the radio detected ``ultradense \ion H2 regions'' ({UD\ion
H2}s) have masses greater than $\sim 10^5 M_\odot$.  We model the
radio spectral energy distributions as both constant density \ion H2
regions and \ion H2 regions with power-law electron density gradients.
These models suggest the {UD\ion H2}s have radii ranging between $\sim
2 - 4$~pc and average electron densities of $\sim 10^3 -
10^4$~cm$^{-3}$ (with peak electron densities reaching values of $\sim
10^5 - 10^6$~cm$^{-3}$).  The pressures implied by these densities are
$P/k_B \sim 10^7 - 10^{10}$~cm$^{-3}$~K, several orders of magnitude
higher than typical pressures in the Galactic ISM.  The inferred \ion
H2 masses in the {UD\ion H2}s are $\sim 2 - 8 \times 10^3 M_\odot$;
these values are $<5$\% of the embedded stellar masses, and anonamously
low when compared to optically visible young clusters.  We suggest
that these low \ion H2 mass fractions may be a result of the extreme
youth of these objects.

\end{abstract}


\section{INTRODUCTION}
The study of ``super star clusters'' (SSCs) has been revolutionized by
the availability of high spatial resolution optical observations, such
as those obtained with the Hubble Space Telescope (HST).  These
clusters appear to be common in starburst and merging galaxy systems
\citep{whitmore02}, and theory suggests that extreme molecular gas
pressures are a prerequisite for their formation
\citep[e.g.][]{elmegreen97,elmegreen02}.  Many SSCs have properties
that are consistent with being adolescent globular clusters
\citep[although the necessary conditions for their survival remain
unclear,][]{gallagher02}.

Compared to globular clusters, SSCs are extremely young objects.
However, from a {\it formation} perspective, everything has already
happened by the time these massive clusters have emerged into optical
light \citep{johnson02}.  Over the last few years, the study of SSCs
has undergone a new revolution with the discovery of ultra-young SSCs
that are still deeply embedded in their birth material.  The pre-natal
clusters have typically been identified in the radio (cm) regime as
compact optically thick free-free sources having turnover frequencies
$\gtrsim 5$~Ghz.  A sample of these objects have now been found in a
number of galaxies \citep[e.g.][]{kj99,turner00, tarchi00, neff00,
johnson01, beck02, plante02, johnson03}.  On a vastly smaller scale,
objects with similar spectral morphologies are associated with
extremely young massive stars in the Milky Way; these objects are
known as ``ultracompact \ion H2 regions'' (UC{\ion H2}s)
\citep[e.g.][]{wc89}.  UC{\ion H2}s result from newly formed massive
stars that are still embedded in their birth material and ionize
compact ($r<<1$~pc) and dense ($n_e\gtrsim 10^4$~cm$^{-3}$) \ion H2
regions within their natal cocoons.  The apparent similarity between
{UC{\ion H2}s and the scaled-up extragalactic radio sources led
\citet{kj99} to dub these extragalactic objects ``ultradense \ion H2
regions'' ({UD\ion H2}s).

Given the limited observations of {UD\ion H2}s that are currently
available, attempts to constrain their physical properties have been
fairly crude.  However, observations have provided estimates for the
physical properties of these objects that are truly remarkable: their
stellar masses can exceed $\sim 10^5 M_\odot$, the radii of their \ion
H2 regions are typically a few parsecs, their electron densities
suggest pressures of $P/k_B\gtrsim 10^8$~K, and they appear to have
ages of less than $\sim 1$~Myr \citep{kj99, vacca02, johnson03}.
Based on these estimates and comparisons to UC{\ion H2}s in the Milky
Way, a physical scenario for these ultra-young SSCs has been developed
as that of compact (yet massive) clusters of stars that ionize
extremely dense \ion H2 regions, which are in turn enveloped by warm
dust cocoons.

The dense \ion H2 regions can be observed in the radio regime, and the
radio spectral energy distributions have turnover frequencies that
increase with increasing electron density. The dust cocoons can be
observed in the infrared to sub-millimeter and may have temperature
profiles similar to individual massive protostars in the Milky Way
\citep{vacca02}, although the current lack of observations in these
wavelength regimes does not allow rigorous constraints to be placed
on the cocoon properties.  For example, it is unclear whether the
constituent stars are surrounded by individual cocoons, or whether the
cocoons have merged and the entire cluster is enveloped in a common
cocoon.  Presumably the relative morphology of the dust cocoon(s) and
the stars depends on both the stellar density and evolutionary state
of the cluster.  Insufficient radio observations also make it
difficult to model the properties of the embedded \ion H2 regions;
while the nature of a compact radio source (non-thermal, thermal, or
optically thick) can be determined from only a pair of high frequency
data points, observations at a number of frequencies are required to
accurately model spectral turnovers and constrain radii and densities.

In order to distinguish between the different physical regions
associated with pre-natal SSCs, throughout this paper we will use the
following nomenclature: ``cluster'' or ``SSC'' refers only to the
embedded {\it stellar} content, ``{UD\ion H2}'' refers to the dense
\ion H2 region surrounding the stellar cluster, and ``cocoon'' refers
to the dust cocoon surrounding the \ion H2 region.

The starburst galaxy Henize~2-10 (He~2-10) is an excellent system in
which to study the properties of ultra-young clusters in detail.  At a
distance of only 9~Mpc \citep[$H_0 = 75$ km s$^{-1}$
Mpc$^{-1}$;][]{vacca92}, it is one of the nearest galaxies known to
host multiple {UD\ion H2}s \citep{kj99} as well as a large number of
SSCs that have already emerged into optical and ultraviolet light
\citep{johnson00, conti94}.  Mid-IR observations obtained by
\citet{vacca02} and \citet{beck01} confirm that the {UD\ion H2}s are
surrounded by warm dust cocoons.  The dust cocoons surrounding the
{UD\ion H2}s in He~2-10 are so luminous that \citet{vacca02} find they
are responsible for at least 60\% of the entire mid-IR flux from this
galaxy.  This percentage of mid-IR flux due to clusters with ages
$\lesssim 1$~Myr old is especially remarkable considering that He~2-10
has been undergoing an intense starburst for several Myr, and hosts
nearly 80 optically visible SSCs \citep{johnson00}.

The original discovery of the {UD\ion H2}s in He~2-10 by \citet{kj99}
utilized observations at only two radio wavelengths in order to
estimate their physical properties.  In this paper we present radio
observations of the {UD\ion H2}s in He~2-10 at five wavelengths (6~cm,
3.6~cm, 2~cm, 1.3~cm, and 0.7~cm) using relatively well-matched
synthesized beams in order to more completely map their radio spectral
energy distributions and better constrain their physical properties.

\section{OBSERVATIONS}

We obtained new Q-band (43~Ghz, 0.7~cm), K-band (22~Ghz, 1.3~cm), and
U-band (15~Ghz, 2~cm) observations of He~2-10 from February 2001 to
January 2003 with the Very Large Array (VLA)
\footnote{The National Radio Astronomy Observatory is a facility of
the National Science Foundation operated under cooperative agreement
by Associated Universities, Inc.}.  VLA archival data at U-band,
X-band (8~Ghz, 3.6~cm), and C-band (5~Ghz, 6~cm) from May 1994 to
January 1996 were also retrieved and re-analyzed.  All of these data
sets are summarized in Table~\ref{observations}.  Based on the scatter
in the VLA Flux Calibrator database, we estimate the resulting flux
density scale at each wavelength is uncertain by $\lesssim 10$\%.

\begin{deluxetable}{lcccccc}
\tabletypesize{\scriptsize}
\tablecaption{VLA Observations of He~2-10\label{observations}}
\tablewidth{0pt}
\tablehead{
\colhead{$\lambda$}&\colhead{Antenna}&\colhead{Date}&\colhead{Obs. Time }
&\colhead{Flux}& \colhead{Phase} & \colhead{Phase Calib.}\\

\colhead{(cm)}&\colhead{Config.} &\colhead{Observed}&\colhead{(hours)}
&\colhead{Calib.}& \colhead{Calib.} & \colhead{$F_\nu$ (Jy)}}
\startdata
0.7 & C-array   & 2003 Jan 05 & 3.7 & 3C286 & 0836-202 & $1.89\pm0.06$ \\
0.7 & C-array   & 2002 Nov 21 & 3.8 & 3C286 & 0836-202 & $2.21\pm0.04$ \\
1.3 & BnA-array & 2001 Feb 04 & 3.1 & 3C286 & 0836-202 & $2.53\pm0.05$ \\
2.0 & BnA-array & 2001 Feb 04 & 0.5 & 3C286 & 0836-202 & $2.65\pm0.09$ \\
2.0 & B-array   & 1996 Jan 03 & 2.5 & 3C286 & 0836-202 & $2.98\pm0.01$ \\
3.6 & B-array   & 1996 Jan 03 & 0.7 & 3C286 & 0836-202 & $3.24\pm0.01$ \\
3.6 & A-array   & 1995 Jun 30 & 0.5 & 3C48  & 0836-202 & $2.41\pm0.06$ \\
3.6 & BnA-array & 1994 May 14 & 0.6 & 3C48  & 0836-202 & $1.90\pm0.01$ \\
6.0 & A-array   & 1995 Jun 30 & 1.6 & 3C48  & 0836-202 & $2.32\pm0.01$ \\
\enddata
\end{deluxetable}

Calibration was carried out in the Astronomical Image Processing
System (AIPS) data reduction package, including gain and phase
calibration.  The data sets at a given wavelength were combined, and
the combined data sets were inverted and cleaned using the task {\sc
imagr}.  While the uv-coverages at each wavelength are not identical,
an attempt was made to obtain relatively well-matched synthesized
beams by varying the weighting used in the imaging process via the
``robust'' parameter (robust$=5$ invokes purely natural weighting,
while a robust$=-5$ invokes purely uniform weighting).  The robust
values used for each wavelength are listed in Table~\ref{imaging}.
The synthesized beams at each wavelength obtained with this method are
relatively well-matched with the exception of the synthesized beam at
1.3~cm; the synthesized beam at 1.3~cm is approximately 1/3 the area
of the beams at the other wavelengths.  Therefore, the relative fluxes
obtained at 1.3~cm should be regarded as lower limits.  In order to
facilitate comparison, the final images were all convolved to the same
synthesized beam of $0.''95 \times 0.''44$ with a position angle of
-3.2 degrees.  Figure~\ref{Q_X} shows the resulting 0.7~cm contours
overlaid on the 3.6~cm grayscale.

\begin{deluxetable}{lcccc}
\tabletypesize{\scriptsize}
\tablecaption{Imaging Parameters \label{imaging}}
\tablewidth{0pt}
\tablehead{
\colhead{$\lambda$}&\colhead{Weighting}&\colhead{Synth. Beam}&\colhead{P.A.}
&\colhead{RMS noise}\\
\colhead{(cm)}&\colhead{(robust value)}&\colhead{($''\times ''$)}
&\colhead{($\degr$)}&\colhead{(mJy/beam)} }

\startdata
0.7 & 1.0 & $0.95\times 0.44$ & -3 & 0.07\\
1.3 & 5.0 & $0.32\times 0.28$ & -3 &0.05\\
2.0 & 1.2 & $0.74\times 0.44$ & -3 & 0.05\\
3.6 & 0.8 & $0.76\times 0.44$ & -3 & 0.03\\
6.0 & 2.0 & $0.95\times 0.44$ & -3 & 0.05\\
\enddata
\end{deluxetable}

\begin{figure}
\centerline{\resizebox{7in}{!}{\includegraphics{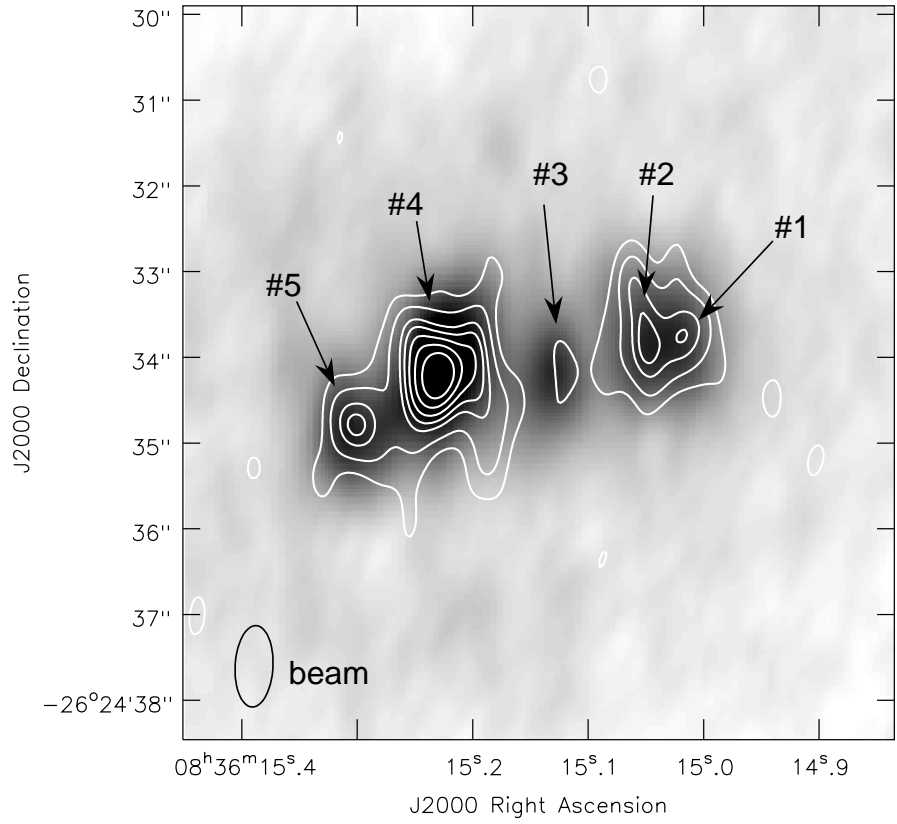}}}
\caption{VLA 0.7~cm contours of He~2-10 overlaid on 3.6~cm
grayscale. The contour levels are $3,5.0,7,8,9,10,11 \times \sigma$
(0.07~mJy/beam).  \label{Q_X}}
\end{figure}

Flux densities of the sources were measured using two methods.  In the
first method, the task {\sc imfit} in AIPS was used to fit the sources
with two-dimensional gaussians; the uncertainty from these results was
estimated by fitting each source using a range of allowed background
estimates and Gaussian profiles.  The second method utilized the {\sc
viewer} program in AIPS++ in order to place apertures around the
sources at each wavelength; several combinations of apertures and
annuli were used in order to estimate the uncertainty in this method.
Both of these methods are fairly sensitive to the background level
that is adopted, and the uncertainty in the resulting flux densities
are dominated by this effect.  The resulting flux densities and their
uncertainties are listed in Table~\ref{fluxes}.  With the exception of
source \#4, all of the 2~cm and 6~cm flux densities reported in this
paper agree within uncertainty to the values presented by
\citet{kj99}; we attribute the single discrepant case to the different
synthesized beams and background estimates used in each analysis.

\begin{deluxetable}{lccccc}
\tabletypesize{\scriptsize}
\tablecaption{Radio Flux Densities of He~2-10 Sources \label{fluxes}}
\tablewidth{0pt}
\tablehead{
& \colhead{$F_{0.7cm}$}& \colhead{$F_{1.3cm}$} 
& \colhead{$F_{2cm}$}& \colhead{$F_{3.6cm}$}& \colhead{$F_{6cm}$}\\
\colhead{Source} &\colhead{(mJy)}&\colhead{(mJy)}&\colhead{(mJy)}
&\colhead{(mJy)}&\colhead{(mJy)}
}
\startdata
1&$0.79\pm0.13$&$0.66\pm0.15$&$0.91\pm0.12$&$0.70\pm0.14$&$0.61\pm0.18$\\
2&$1.08\pm0.14$&$0.70\pm0.19$&$0.97\pm0.14$&$0.87\pm0.19$&$0.66\pm0.17$\\
3&$0.23\pm0.05$&$0.43\pm0.15$&$0.68\pm0.13$&$0.78\pm0.22$&$0.66\pm0.18$\\
4&$2.91\pm0.57$&$2.36\pm0.48$&$3.11\pm0.38$&$3.03\pm0.52$&$2.26\pm0.36$\\
5&$0.89\pm0.21$&$0.65\pm0.16$&$0.83\pm0.15$&$0.76\pm0.10$&$0.58\pm0.07$\\
\enddata
\end{deluxetable}

\section{PROPERTIES OF THE UD{\ion H2}s}
\subsection{Production Rate of Ionizing Photons \label{Q}}
In order to estimate the embedded stellar content of the {UD\ion H2}s in
He~2-10, their radio luminosities can be used to determine the production
rate of Lyman continuum photons.  Following \citet{condon92},
\begin{equation}
{{Q_{Lyc}}}
\geq6.3\times10^{52}~{{\rm s}^{-1}}\Big({{T_e}\over{10^4~{\rm K}}}\Big)^{-0.45}
\Big({{\nu}\over{{\rm Ghz}}}\Big)^{0.1}
\Big({{L_{thermal}}\over{10^{27}~{\rm erg~s^{-1}~Hz}^{-1}}}\Big).
\end{equation}
In the application of this equation (which assumes the emission is
purely thermal and optically thin), it is advantageous to use
measurements made at the highest radio frequency available for two
reasons: (1) the higher the frequency, the less likely it is to
contain a significant amount of non-thermal contaminating flux, and
(2) the higher frequency emission suffers from less self-absorption and
is therefore more likely to be optically thin.  For these reasons, we
use the 43~Ghz (0.7~cm) observations to determine $Q_{Lyc}$ for the
{UD\ion H2}s in He~2-10.  An electron temperature must also be assumed,
and we adopt a ``typical'' \ion H2 region temperature of $T_e =
10^4$~K; the uncertainty in $Q_{Lyc}$ due to this assumption is
$\lesssim 20$\%.  The resulting $Q_{Lyc}$ values determined using this
method are listed in Table~\ref{properties}, and they range from
$Q_{Lyc} \approx 7 \times 10^{51} - 26 \times 10^{51}$~s$^{-1}$.  To
put these values in context, a ``typical'' O-star \citep[hereafter O*,
the equivalent to type O7.5V;][]{vacca94,vacca96} has an ionizing flux
of $Q_{Lyc}=1 \times 10^{49}$~s$^{-1}$.  Therefore, the $Q_{Lyc}$ values
determined for the {UD\ion H2}s in He~2-10 imply the equivalent of 700
- 2600 O*-stars in each cluster.  

\begin{deluxetable}{lcccccccccc}
\tabletypesize{\scriptsize}
\tablecaption{Estimated Properties of the {UD\ion H2}s 
\label{properties}}
\tablewidth{0pt}
\tablehead{
\colhead{} &  \multicolumn{2}{c}{} & 
\multicolumn{3}{c}{Constant Density Models} & \colhead{} &
\multicolumn{4}{c}{Density Gradient Models} \\
\cline{4-6} \cline{8-11} \\ 

& \colhead{$Q_{Lyc}$\tablenotemark{a}} & \colhead{M$_{stars}$} & 
\colhead{r} & \colhead{$n_e$} & \colhead{M$_{\rm HII}$} && 

& \colhead{r$_{1/2}$} & \colhead{$n_{1/2}$} & 
\colhead{M$_{\rm HII}$}\\

\colhead{\#}&\colhead{($\times 10^{51}$s$^{-1}$)} & 
\colhead{($\times 10^5 M_\odot$)} 
&\colhead{(pc)} & \colhead{($\times 10^3$cm$^{-3}$)} 
&\colhead{($\times 10^3 M_\odot$)}  & &\colhead{$\gamma$}  & \colhead{(pc)} & 
\colhead{($\times 10^3$cm$^{-3}$)} & \colhead{($\times 10^3 M_\odot$)}
\
}
\startdata
1 & 7.1 & 1.2  & $1.8^{+1.1}_{-0.4}$ & $6.7^{+3.2}_{-3.7}$ & 4.0 & &
-0.9 & 1.7 & 3.6 & 5.5\\
\\
2 & 9.7 & 1.6  & $1.8^{+0.7}_{-0.4}$ & $7.3^{+3.7}_{-3.0}$ & 4.4 & &
-1.3 & 2.1 & 2.4 & 7.9\\
\\
4 & 25.9& 4.3  & $3.9^{+1.2}_{-0.9}$ & $3.9^{+2.1}_{-1.3}$ & 2.4 & &
-0.0 & 3.1 & 3.9 & 2.4\\
\\
5 & 7.9 & 1.3  & $1.7^{+0.4}_{-0.2}$ & $7.3^{+1.6}_{-2.1}$ & 3.7 & &
-1.1 & 1.9 & 2.9 & 6.2\\ 

\enddata \tablenotetext{a}{$Q_{Lyc}$ values as determined from the
0.7~cm flux densities, assuming an HII region temperature of
$10^4$~K.}
\end{deluxetable}

There is an important caveat in regard to the $Q_{Lyc}$ values
discussed above.  A number of Galactic {UC\ion H2}s are observed to be
associated with diffuse extended emission.  For example, in the
\citet{kim01} sample of 16 {UC\ion H2}s, all of the objects were
associated with extended radio recombination line emission.  Likewise,
\citet{kurtz99} find that 12 out of 15 {UC\ion H2}s in their sample
are associated with extended emission.  These studies conclude that
typically $\gtrsim 80$\% of the ionizing flux from the embedded stars
in {UC\ion H2}s is escaping to the outer envelope, possible due to a
clumpy density structure within the parent molecular cloud.  In the
case of He~2-10, there also appears to be a somewhat diffuse thermal
background in the immediate vicinity of the {UD\ion H2}s.  If {UD\ion
H2}s are leaking ionizing flux in similar proportions to their
Galactic counterparts, the inferred stellar content from the $Q_{Lyc}$
values determined above may be a significant underestimate.  To
estimate an upper limit on the possible leakage of ionizing flux from
the {UD\ion H2}s, the ionizing flux was measured from the entire
region surrounding the {UD\ion H2}s.  The entire region has an
ionizing flux of $Q_{Lyc}\sim 8\times 10^{52}$~s$^{-1}$, approximately
40\% higher than the sum of the ionizing fluxes from the individual
{UD\ion H2}s.  This ``extra'' flux is only an upper limit on the
ionizing flux that could be escaping from the {UD\ion H2}s as it could
also be due to contributions from other more evolved clusters in the
region.

The masses of the embedded stellar clusters can be estimated from
their Lyman continuum fluxes using the Starburst99 models of
\citep{leitherer99}.  For a cluster less than 1~Myr old formed in an
instantaneous burst with a Salpeter IMF, 100~$M_\odot$ upper cutoff,
1~$M_\odot$ lower cutoff (note that reducing the lower mass cutoff to
0.1~$M_\odot$ increases the cluster mass by a factor of $\sim 2.5$),
and solar metallicity, a $10^6 M_\odot$ cluster has $Q_{Lyc} \approx 6
\times 10^{52}$~s$^{-1}$.  Assuming that $Q_{Lyc}$ scales directly
with the cluster mass, we find that the {UD\ion H2}s in He~2-10 are
powered by stellar clusters with masses $\sim 1.1 - 4.3 \times 10^5
M_\odot$ (Table~\ref{properties}).  The mass derived for source~4
from these radio observations is in good agreement with, but a factor
of 2 lower than the mass derived by \citet{vacca02} using the
bolometric luminosity of the cluster (after taking into account the
0.1~$M_\odot$ lower mass cutoff used by Vacca et al.).  While the
small difference between the mass derived from the radio and
bolometric luminosities is certainly contained within the
uncertainties, the direction of the difference possibly suggests that
a fraction of the ionizing photons are being absorbed by dust within
the {UD\ion H2}.  Given the possible leakage of ionizing photons from
the {UD\ion H2}s and/or their absorption by dust, these mass estimates
derived from the radio luminosities are lower limits, and the cluster
masses could be as much as $\sim 10 \times$ higher than these values.

\subsection{Comparison to Model \ion H2 Regions \label{models}}
In order to gain additional physical insight about these \ion H2
regions, we can invoke simple models.  Following the models of
\citet{johnson03}, we model the compact radio sources as spherical
\ion H2 regions with constant electron temperature of $10^4$~K under
two different assumptions: (1) the electron density profile is
constant, and (2) the electron density profile is allowed to vary as a
power-law of the form $n_e \propto r^\gamma$.  For both sets of
models, the flux density measurements at 43~Ghz, 15~Ghz, 8~Ghz, and
5~Ghz were used to fit the models (in a least-squared sense); the
22~Ghz data points were excluded from the models fits as they are only
lower limits.

The major difference between these models and those used by
\citet{kj99} and \citet{johnson01} is the allowance for a power-law
density profile within a given \ion H2 region.  We also include
limb-darkening, which has an effect in the optically-thin regime at
high frequencies.  However, the results we obtain with these new
models are in excellent agreement with the results from the models
used in the above papers; this provides additional confidence in the
robustness of the model results.  However, the additional radio data
points presented in this paper allow us to place more rigorous
constraints on the model parameters than \citet{kj99}.

For the models assuming constant electron density, the data are best
fit by \ion H2 regions with radii that range between r$\approx 2-4$~pc
and electron densities of $n_e \approx 4 - 7 \times 10^3$~cm$^{-3}$.
These best fit models are shown with the dashed line in
Figure~\ref{mod_figs}.  The modeled radii and densities were then used
to determine total \ion H2 masses of the {UD\ion H2}s of $\sim 4 - 24
\times 10^3 M_\odot$.  Given that we have two free parameters ($n_e$
and r) and four data points, the models are over-constrained and we
determine the possible range in the model parameters by finding
the minimum and maximum values for $n_e$ and r that can fit the data
within uncertainty.  All of these properties are listed in
Table~\ref{properties}.

\begin{figure}
\centerline{\resizebox{3in}{!}{\includegraphics{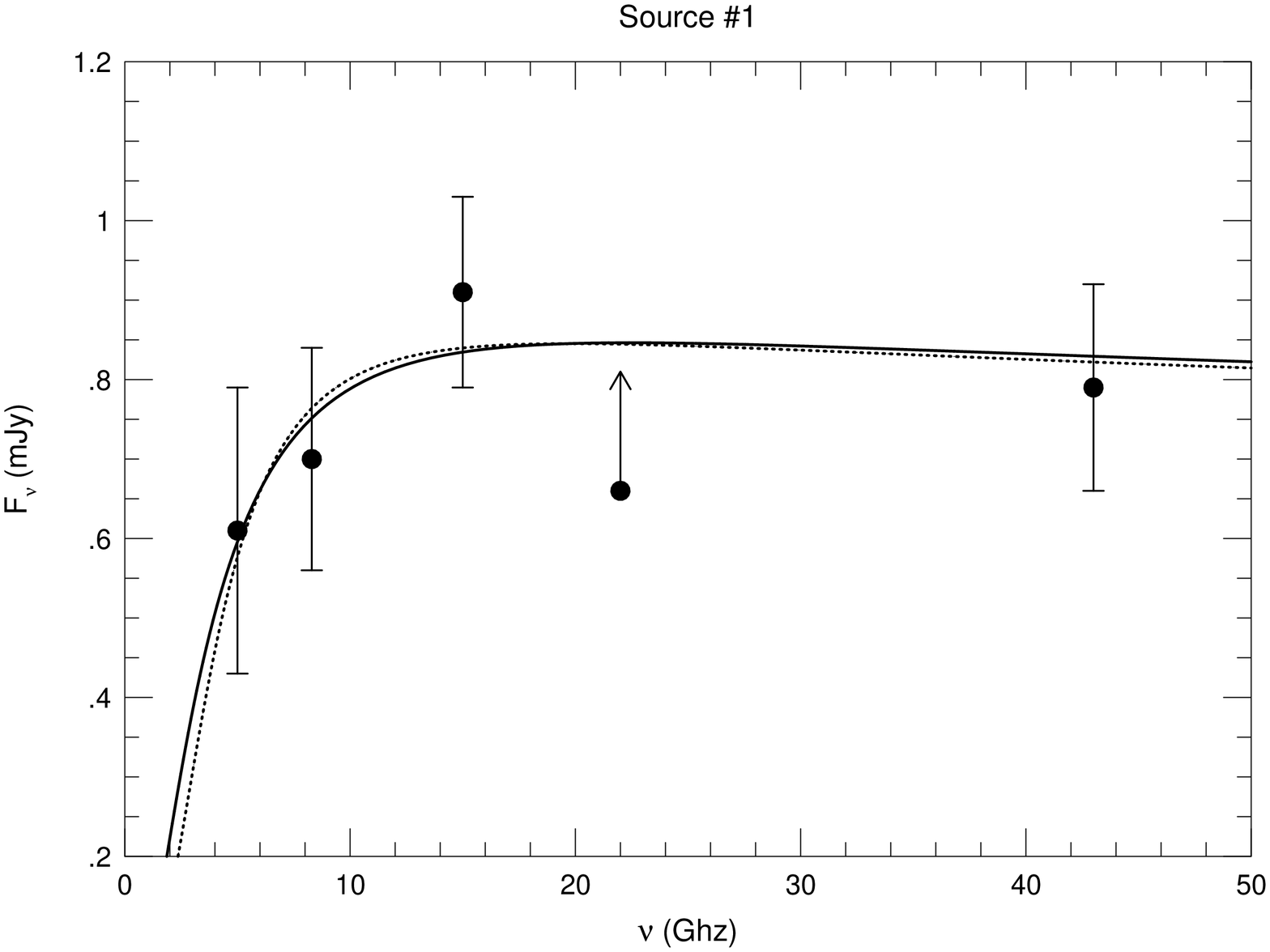}}
\resizebox{3in}{!}{\includegraphics{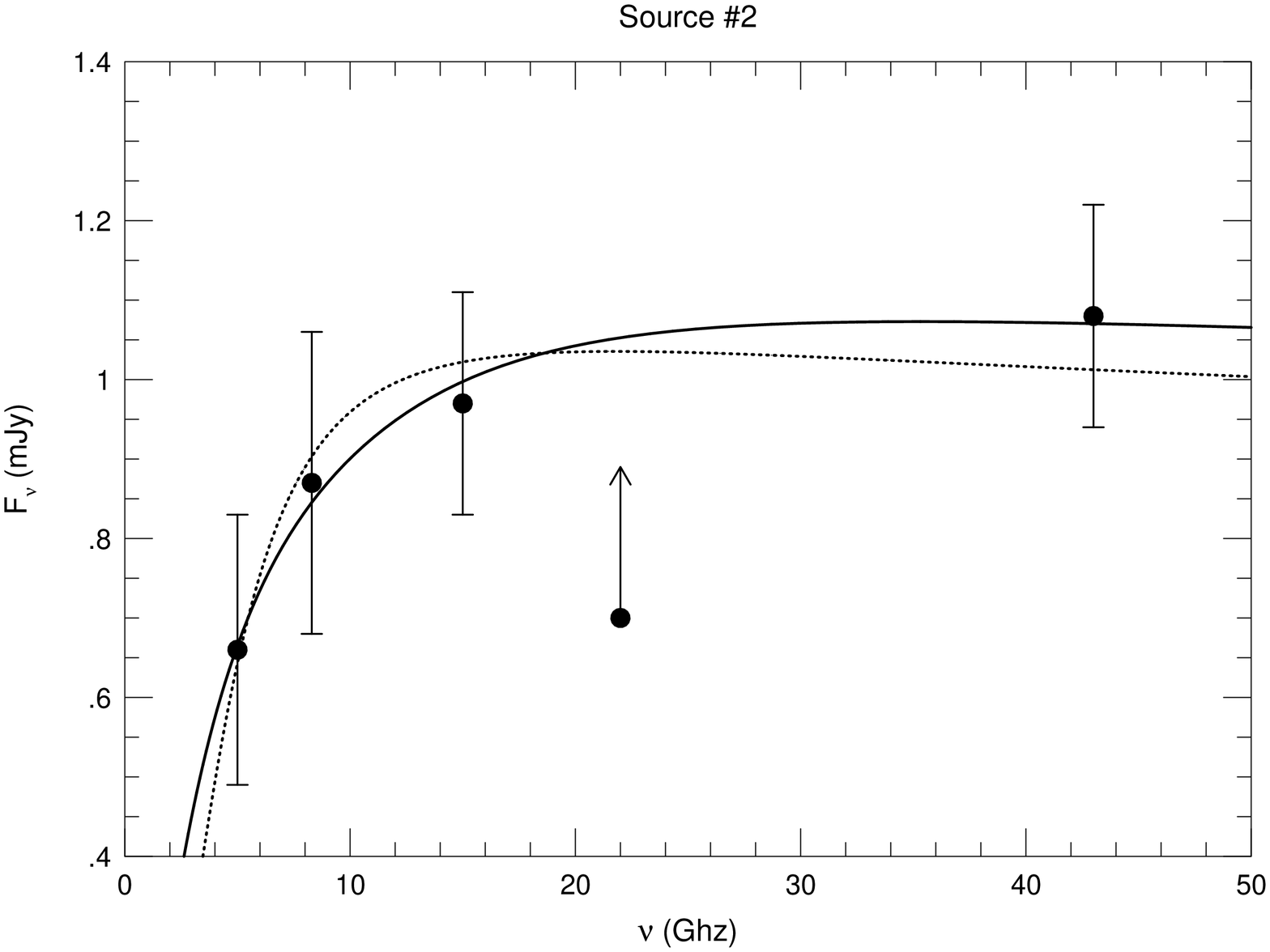}}}
\centerline{\resizebox{3in}{!}{\includegraphics{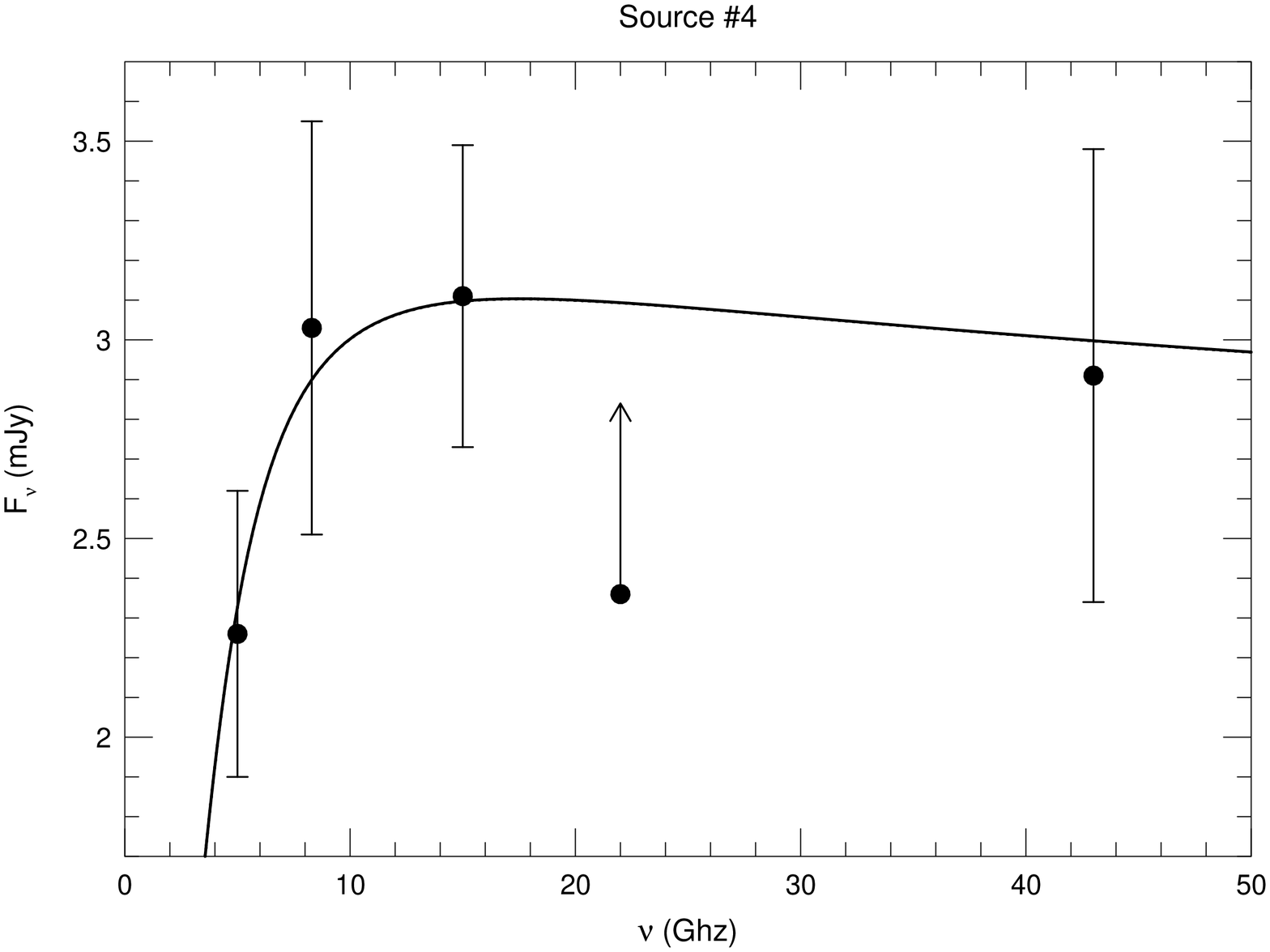}}
\resizebox{3in}{!}{\includegraphics{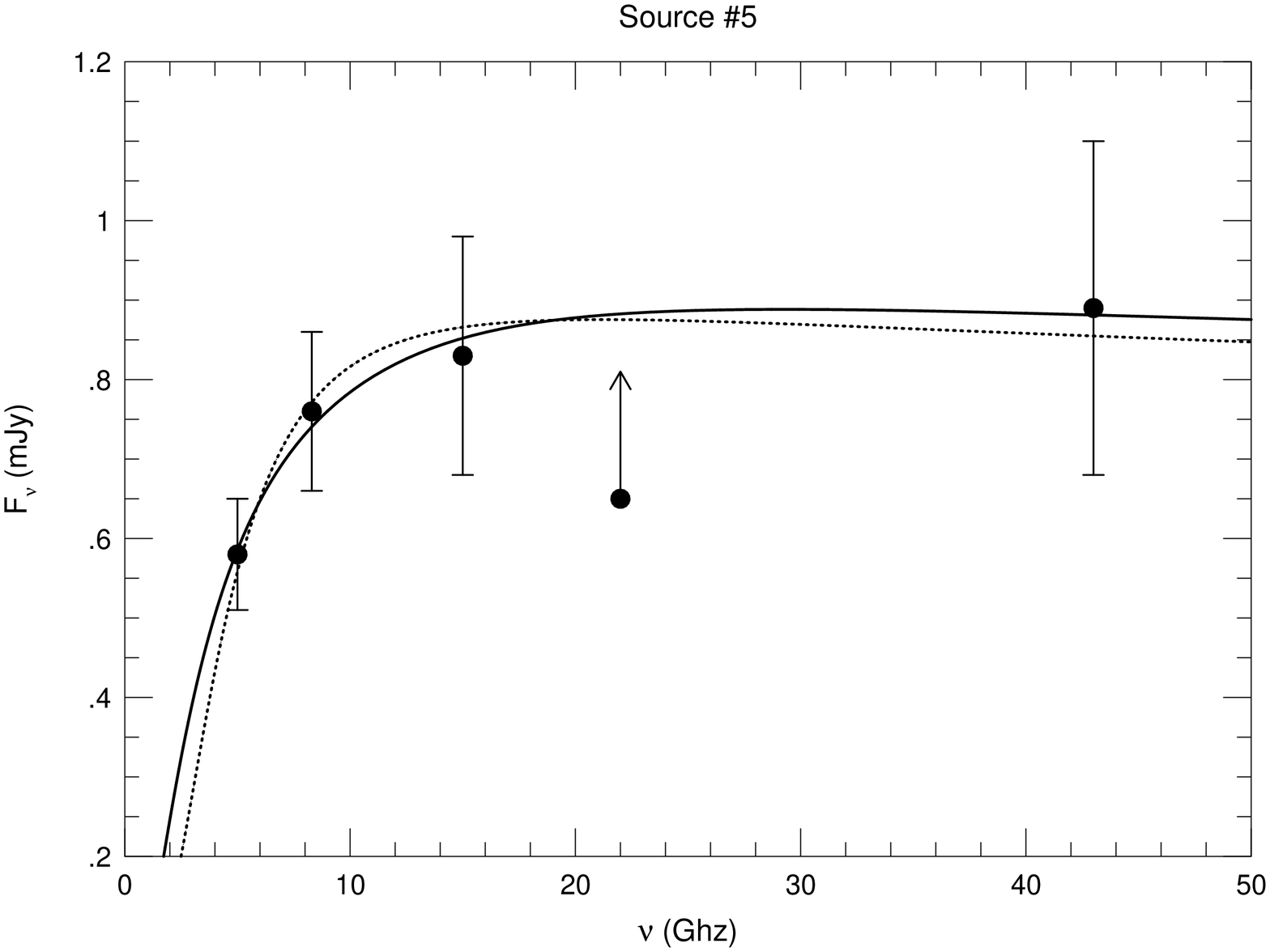}}}
\caption{\ion H2 region model fits to the {UD\ion H2}s in He~2-10
(source 3 is not included as its spectral energy distribution cannot
be fit by a thermal model).  The dashed lines are the best fits from
the constant density models, and the solid lines are the best fits
from the models allowing for a density gradient.  Because the
synthesized beam at 22~Ghz is significantly smaller than the beams at
the other wavelengths, the flux density measured at 22~Ghz is a lower
limit and was not used to fit the models. \label{mod_figs}}
\end{figure}

The second set of models allowed for a density gradient of the form
$n_e \propto r^\gamma$.  The main affect of such a gradient on the
modeled spectral energy distributions is to make the turnover less
abrupt.  For sources 1, 2, and 5 these models produce a marginally
better fit than the constant density models; the best fit model for
source 4 had $\gamma = 0$, which is equivalent to the constant density
model.  These best fit models are shown with a solid line in
Figure~\ref{mod_figs}.  For these models, the best fit density
gradient ($\gamma$), the half-mass radius (r$_{1/2}$), the electron
density at the half-mass radius ($n_{1/2}$), and the total inferred
\ion H2 mass (M$_{HII}$) are listed in Table~\ref{properties}.  For
the sources best fit with a density gradient, the electron densities
exceed values of $\approx 10^4$~cm$^{-3}$ in the inner 0.5~pc, and the
core densities can reach values close to $\sim 10^6$~cm$^{-3}$.

The inferred \ion H2 mass for each {UD\ion H2} is $< 5$\% of the
embedded stellar mass ($\sim 3$\% -5\% for sources 1, 2, and 5, and
$\sim 1$\% for source 4), which is a strikingly low value when
compared to optically visible young clusters.  For example, the R136
cluster in the 30 Doradus nebula has a stellar mass of $\sim 6 \times
10^4 M_\odot$ \citep{hunter95}, but the \ion H2 associated with the
region has a mass of $\sim 1 \times 10^5$ \citep{peck97}.  As the
radio observations presented here do not suffer from extinction, it is
not likely that the low \ion H2 mass fraction for these objects is due
to intervening absorption.  Instead, we suggest that the low \ion H2
mass fraction might be due to the extreme youth of the stellar
clusters; much of the gas mass in the vicinity of the cluster may
still be shielded from the ionizing radiation, and therefore still in
molecular or neutral atomic form.

The densities suggested by the models described above imply
tremendously high pressures of $P/k_B \sim 10^7 - 10^{10}$~cm$^{-3}$~K
within the {UD\ion H2}s.  Pressures of this magnitude are also typical
in Galactic {UC\ion H2}s \citep[e.g.][and references
therein]{churchwell99}, but they are extremely high compared to
typical pressures in the ISM of $P/k_B \sim 10^3 - 10^4 $~cm$^{-3}$~K
\citep{jenkins83}.  These high pressures may be one of the
requirements for the formation of SSCs \citep[e.g.][]{elmegreen97,
elmegreen02}.  However, there are two caveats to bare in mind when
considering these estimated pressures: (1) A cluster itself will
contribute to the pressure of the surrounding medium.  We can estimate
the affect of the cluster by assuming the natal \ion H1 cloud had a
gas temperature of $T_{HI}\sim 10^2 K$, and the photoionization of the
cluster increases the temperature to $T_{HII}\sim 10^4 K$; therefore,
the cluster itself will cause the pressure to increase a factor of
$\sim 10^2$ over the initial value in the natal cloud.  (2) We are not
directly measuring the initial pressure of the newborn \ion H2 region.
Despite the youth of these objects, the \ion H2 regions have had some
time to expand toward pressure equilibrium.  Consequently, the
pressures inferred at this stage in their evolution are {\it lower}
than the initial values.  If an \ion H2 region expands at a sound
speed of roughly $\sim 10$~km/s, it will have expanded from $\sim
1$~pc to $\sim 10$~pc in 1~Myr, causing the pressure to decrease by
roughly a factor of $\sim 10^2$.  These are fairly crude estimates,
however even in the case that effect (1) is dominant, the pressures
implied for the quiescent natal material are $P/k_B \sim 10^5 -
10^{8}$~cm$^{-3}$~K.

Source 3 cannot be fit by any purely thermal model
(Figure~\ref{source3}).  Although this source appears to have a slight
turnover around 5~Ghz, the data at the remaining frequencies are fit
by a fairly steep negative power-law of $\alpha \sim -1$ (where $S_\nu
\propto \nu^\alpha$), indicating a non-thermal origin.  The nature of
this source is discussed in Vacca et al. (in prep).

\begin{figure}
\plotone{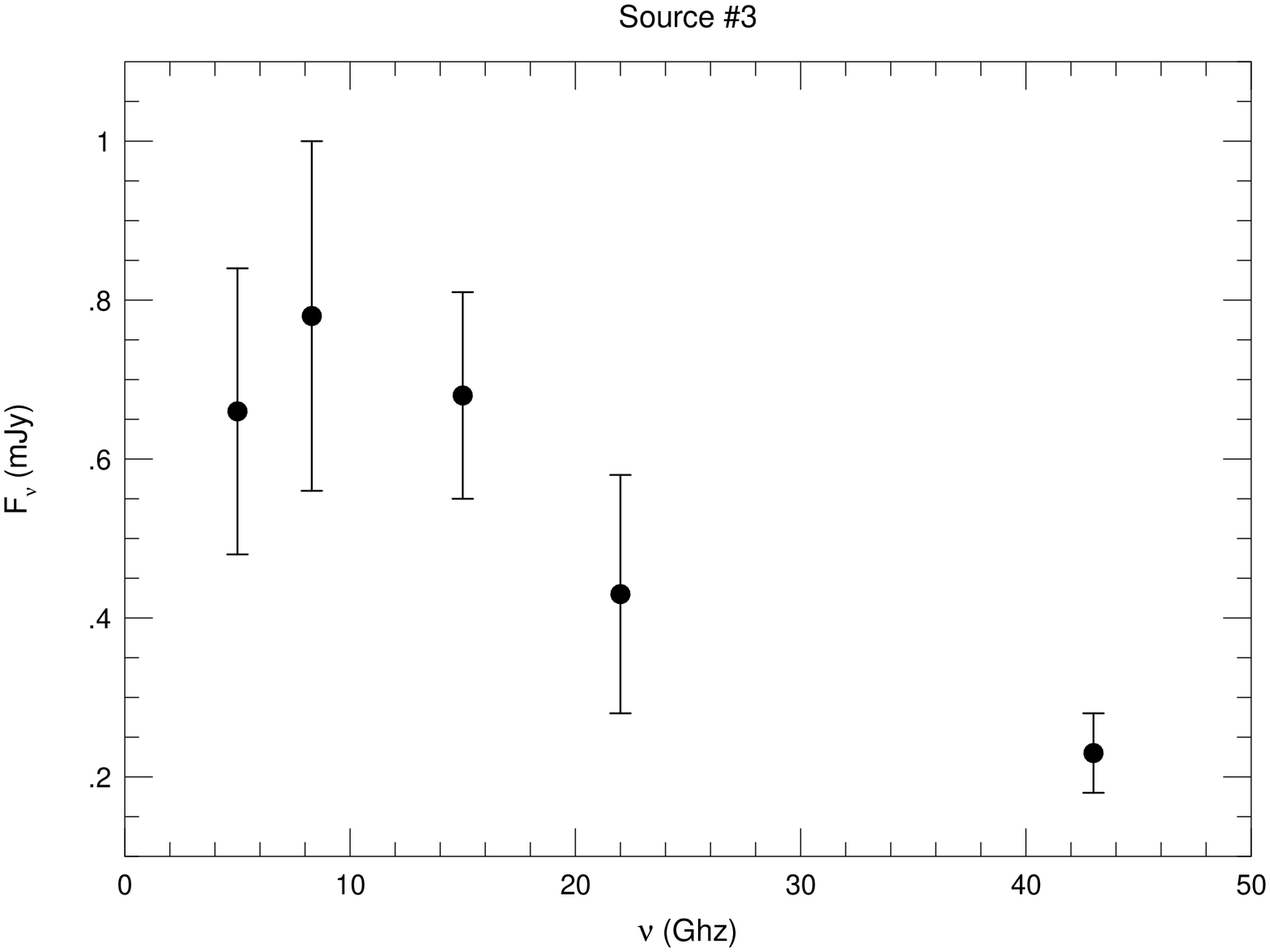}
\caption{The spectral energy distribution of source 3.  This spectral
energy distribution cannot be fit by a thermal model. 
\notetoeditor{Please reproduce Figure 3 at approximately the same
scale as the individual figures in Figure 2.} \label{source3}}
\end{figure}

\section{SUMMARY}
Multi-frequency radio observations of the deeply embedded massive star
forming regions in He~2-10 have allowed us to better constrain their
physical properties.  The ionizing fluxes calculated from the thermal
radio continuum have values ranging from $Q_{Lyc} \approx 7 \times
10^{51} - 26 \times 10^{51}$~s$^{-1}$, which imply the clusters have
stellar masses $> 10^5 M_\odot$.  These masses are fully consistent
with the masses estimated for the optically visible SSCs by
\citet{johnson00}.  We model the \ion H2 regions under the assumptions
of both constant electron densities and a power-law electron density
gradients.  These models suggest that the dense \ion H2 regions have
radii $< 4$~pc (and as small as 1.7~pc), and the electron densities
may reach values as high as $\sim 10^6$~cm$^{-3}$ in the cores of
these regions.  The inferred \ion H2 masses for the {UD\ion H2}s are
$< 5$\% of the embedded stellar masses, anomalously low compared to
optically visible young clusters, and possibly due to the extreme
youth of these objects.  The densities derived from the models imply
pressures of $P/k_B \sim 10^7 - 10^{10}$~cm$^{-3}$~K within the
{UD\ion H2}s, typical of Galactic {UC\ion H2}s, and provide us with
observational confirmation of the extremely high pressures involved in
the early stages of super star cluster evolution.

\acknowledgments 

We are grateful the NRAO staff for their assistance, and Claire
Chandler in particular for her advice on the high frequency
observations.  We thank Bill Vacca, Miller Goss, and Jay Gallagher for
many useful discussions on this subject and their comments on the
manuscript.  Remy Indebetouw kindly provided assistance developing the
physical models presented in this paper.  We also thank the anonymous
referee for providing stimulating comments that led to improvements in
the manuscript.  K.E.J. gratefully acknowledges support for this paper
provided by NSF through an Astronomy and Astrophysics Postdoctoral
Fellowship.

\end{document}